\documentclass[showpacs,onecolumn,amsfonts,amsmath,amssymb,pre,aps,dvips,superscriptaddress,preprint]{revtex4}

\usepackage{graphicx,epsfig}
\usepackage{ae}
\usepackage{bm}
\usepackage[T1]{fontenc}

\begin{document}

\title{Single-file and normal diffusion of magnetic colloids in modulated channels}

\author{D.~Lucena}
\email{diego@fisica.ufc.br}
\affiliation{Departamento de F\'isica, Universidade Federal do Cear\'a, Caixa Postal 6030, Campus do Pici, 60455-760 Fortaleza, Cear\'a, Brazil}
\affiliation{Department of Physics, University of Antwerp, Groenenborgerlaan 171, B-2020 Antwerp, Belgium}

\author{J.~E.~Galv\'{a}n-Moya}
\affiliation{Department of Physics, University of Antwerp, Groenenborgerlaan 171, B-2020 Antwerp, Belgium}

\author{W.~P.~Ferreira}
\affiliation{Departamento de F\'isica, Universidade Federal do Cear\'a, Caixa Postal 6030, Campus do Pici, 60455-760 Fortaleza, Cear\'a, Brazil}

\author{F.~M.~Peeters}
\email{francois.peeters@uantwerpen.be}
\affiliation{Department of Physics, University of Antwerp, Groenenborgerlaan 171, B-2020 Antwerp, Belgium}
\affiliation{Departamento de F\'isica, Universidade Federal do Cear\'a, Caixa Postal 6030, Campus do Pici, 60455-760 Fortaleza, Cear\'a, Brazil}

%\date{\today}

\begin{abstract}
Diffusive properties of interacting magnetic dipoles confined in a parabolic narrow channel and in the presence of a periodic modulated (corrugated) potential along the unconfined direction are studied using Brownian dynamics simulations. We compare our simulation results with the analytical result for the effective diffusion coefficient of a single-particle by Festa and d'Agliano [Physica A \textbf{90}, 229 (1978)] and show the importance of inter-particle interaction on the diffusion process. We present results for the diffusion of magnetic dipoles as a function of linear density, strength of the periodic modulation and commensurability factor.
\end{abstract}


\maketitle

\section{Introduction}
Manipulation and control of magnetic colloidal particles have greatly increased over the last years. Recent advances include fabrication of anisotropic magnetic particles \cite{JACS2008} which can have a wide range of applications, from drug deliver mechanisms \cite{PNASMitra,Dobson} to fabrication of tunable self-assembly colloidal devices \cite{AransonNat2011,Vandewalle2012}. Further examples of applications of anisotropic particles are the so-called colloidal molecules \cite{BGates,ZacaNat}, the patchy colloids \cite{JonesNat,XMaoPRE,XMaoNatMat} and the magnetic Janus colloids \cite{JYanNat}. The use of magnetic dipoles is particularly interesting due to the fact that the inter-particle interaction potential introduces a natural source of anisotropy. This is achieved by the application of a tunable external static homogeneous \cite{Kantorovich,JFaraudo} or oscillating \cite{SabineKlappPRE,PTiernoPRL2010} magnetic field ($\bm{B}$). Diffusion and transport of colloidal particles in periodic modulated (corrugated) channels \cite{PTiernoPRL2012} represent important phenomena which allows the understanding of several mechanisms in soft condensed matter, e.g., molecular and cell crowding in biological systems \cite{SaxtonBioCoxPRL}, pinning-depinning transition of vortices in type-II superconductors \cite{SlavaPRB2009}, and elastic strings \cite{DongPRL93WiesePRL2001}. Theoretical models which describe the trapping dynamics of modulated systems include, for instance, continuous time random walk (CTRW) \cite{KlafterAndSilbeyCTRW} and random walk with barriers \cite{HelpernNatPhys2011}. Experimentally, corrugated periodic \cite{EgelhaafSM2011} or random \cite{EgelhaafSM2012} landscapes can be realised, e.g., by light fields allowing the control of the colloidal particles. Furthermore, diffusion in modulated landscapes is often anomalous, i.e., the mean-square displacement $W(t)$ (MSD) follows a power-law [$W(t) \propto t^{\alpha}$] with an exponent $0 < \alpha < 1$ \cite{KlafterPRL2013}.

Diffusion in very narrow channels is governed by single-file diffusion (SFD) \cite{MarchePRL2006}. An interesting quantity in this case is the single-file mobility factor, $F$. This factor has been previously analysed by Herrera-Velarde and Casta\~{n}eda-Priego \cite{CastanedaPRE2008} for the case of a system of repulsive interacting superparamagnetic colloids. In our case, however, the attractive part of the inter-particle interaction potential [Eq.~(\ref{eqIntPotential})] introduces an anisotropy in the system. This means that the external magnetic field, regulated by the magnitude of $\bm{B}$ and its direction $\phi$, now plays an important role in tuning the diffusive properties of the system. The effects of these two parameters in a system without external modulation has been recently investigated by us in Ref.~\cite{LucenaPREMag}. Here we extend these results to the case where the narrow channel is periodically modulated along the unconfined direction. We find that the commensurability between the inter-particle distance and the period of the modulation is an essential factor that strongly influences the diffusion.

This paper is organized as follows. In Sec.~\ref{singlePart} we analyse the simplest one-dimensional (1D) case of a single particle diffusing in the presence of a periodic modulation. The effect of many-particle interactions is analysed in Sec.~\ref{InterMagDipoles} where we present a model of interacting magnetic dipoles. The equations used to describe the dynamics of the particles, together with definitions and simulation parameters are given. In Sec.~\ref{MSDSecDensity} we fix the linear density of the system, and analyse the diffusion coefficient and the single-file mobility factor as a function of the strength of the periodic modulation. Sec.~\ref{linDens} is dedicated to the analysis of the influence of the linear density on diffusion and in Sec.~\ref{CommSection} we study the influence of the commensurability factor on diffusion. The case of moderate density is investigated in Sec.~\ref{AnisoSFD} where the diffusion process becomes anisotropic due to the competition between the periodic modulation and the transversal confinement. We show that a transversal SFD and a sub-diffusive regime can be induced by tuning the external periodic modulation. The conclusions are presented in Sec.~\ref{ConcDiss}.

\section{Single-particle in an external periodic potential}\label{singlePart}
First, we consider the simplest case of a single-particle diffusing in one dimension and subjected both to Brownian motion and to an external periodic potential landscape. The equation of motion for the particle is given by the overdamped Langevin equation \cite{Risken}:
\begin{equation}
\gamma \frac{dx}{dt} = - \frac{\partial V(x)}{\partial x} + \xi(t),
\end{equation}
where $\gamma$ is the viscosity of the medium, $x$ is the position of the particle, $t$ is time, $V(x)$ is the external one-dimensional periodic potential of the form $V(x) = V_{0}\cos(2 \pi x/L)$, where $V_{0}$ and $L$ are the magnitude and periodicity of the external potential, respectively. Essential here is that $V(x)$ is periodic but it does not necessarily need to be of cosine form. The only condition is that the external potential obeys the periodicity relation, $V(x) = V(x+L)$. Furthermore, $\xi(t)$ is a delta correlated noise which follows the well-known properties (i) $\langle \xi(t) \rangle = 0$ and (ii) $\langle \xi(t)\xi(t') \rangle = 2\gamma k_{B} T \delta(t-t')$. $k_{B}$ is the Boltzmann constant and $T$ is the absolute temperature of the heat bath.

\begin{figure}[ht]
\begin{center}
%\hspace*{-0.2cm}
\includegraphics[width=8.6cm]{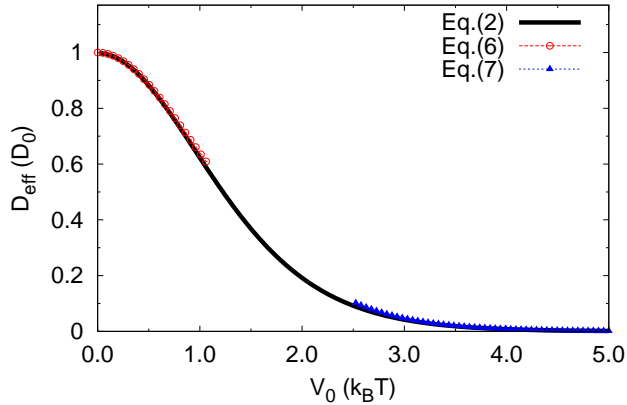}
%\vspace*{-1.2cm}
\end{center}
\caption{Effective self-diffusion coefficient $D_{\text{eff}}/D_{0}$ of a single-particle in one dimension in the presence of a thermal bath and a periodic potential $V(x') = V_{0}\cos(x')$.}\label{DeffFig}
\end{figure}

In the case where the particle is free, i.e., $V_{0} = 0$, it is straightforward to show \cite{Risken} that the self-diffusion coefficient of the particle is given by the Einstein relation $D_{0} = k_{B}T/\gamma$. In the presence of a periodic potential $V(x)$, previous studies \cite{LifsonJChemPhys,FestaPhysA,ReimannPRE} showed that the self-diffusion coefficient of the particle is modified into:
\begin{equation}\label{Deff}
\frac{D_{\text{eff}}}{D_{0}} = \frac{L^{2}}{\int_{0}^{L} dx \exp\{V(x)/k_{B}T\} \int_{0}^{L} dz \exp\{-V(z)/k_{B}T\}}.
\end{equation}
It is easy to see that when $V(x) = 0$, Eq.~(\ref{Deff}) reduces to $D_{0}$ as it should be. If we consider the case of $L=2\pi\sigma$ and $x \rightarrow x'\sigma$ [$V(x') = V_{0}\cos(x')$], the solutions of the integrals in (\ref{Deff}) are known \cite{WatsonBess} and given by
\begin{eqnarray}
\sigma \int_{0}^{2\pi} dx' \exp\{V(x')/k_{B}T\} &=& 2\pi\sigma I_{0}(V_{0}/k_{B}T)\label{integral1},\\
\sigma \int_{0}^{2\pi} dx' \exp\{-V(x')/k_{B}T\} &=& 2\pi\sigma I_{0}(-V_{0}/k_{B}T),\label{integral2}
\end{eqnarray}
where $I_{0}(y)$ are the modified Bessel functions of the first kind and $\sigma$ is a unit of distance. Therefore, the self-diffusion coefficient $D_{\text{eff}}$ depends only on the ratio $V_{0}/k_{B}T$. A series representation of $I_{0}(y)$ can be written as \cite{Gradshteyn}
\begin{equation}\label{BesselApp1}
I_{0}(y) = 1 + \frac{(y/2)^{2}}{(1!)^{2}} + \frac{(y/2)^{4}}{(2!)^{2}} + \ldots
\end{equation}
Taking the first order approximation in Eq.~(\ref{BesselApp1}), we have that $D_{\text{eff}}/D_{0}$ is given by
\begin{equation}\label{ylessthan1}
\frac{D_{\text{eff}}}{D_{0}} \simeq \frac{1}{[1 + (y/2)^2]^2}.
\end{equation}
Note that for $y = V_{0}/k_{B}T \ll 1$, $D_{\text{eff}}/D_{0} \rightarrow 1$, as expected. On the other hand, for $y = V_{0}/k_{B}T \gg 1$, the modified Bessel function $I_{0}(y)$ can be written to a first order approximation as \cite{AbramoHandbook} $I_{0}(y) \simeq e^{y}/\sqrt{2\pi y}$. Therefore, $D_{\text{eff}}/D_{0}$ has the form
\begin{equation}\label{ygreatthan1}
\frac{D_{\text{eff}}}{D_{0}} \simeq (2\pi y) e^{-2y}.
\end{equation}
For $y = V_{0}/k_{B}T \gg 1$, $D_{\text{eff}}/D_{0} \rightarrow 0$. Both limiting cases (\ref{ylessthan1}) and (\ref{ygreatthan1}) are shown in Fig.~\ref{DeffFig}.

%, and we show this dependence in Fig.~\ref{DeffFig}.

\section{Interacting magnetic dipoles}\label{InterMagDipoles}
We now turn to the problem where instead of a single-particle we have $N$ interacting magnetic dipoles of diameter $\sigma$ and magnetic moment $\bm{\mu}$ diffusing in the plane $(x,y)$. The geometry of the plane is then modulated by two external potentials, namely (i) a parabolic transversal confinement potential in the $y$ direction and (ii) a periodic potential in the $x$ direction. We also apply an external homogeneous magnetic field $\bm{B}$ which can form an angle $0^{o} \leq \phi \leq 90^{o}$ with the $x$ axis. In this more complex situation, the equations of motion which describe the dynamics of particle $i$ are given by $N$ overdamped coupled Langevin equations
\begin{eqnarray}
\gamma \dot{\textbf{r}}_{i} &=& - \sum_{j>i} \bm{\nabla} V^{\text{int}}_{ij} - \bm{\nabla} [V_{\text{mod}}(x_{i}) + V_{\text{conf}}(y_{i})] + \bm{\xi}_{i}(t)\label{eqMotion1},\\
\gamma\sigma^{2} \dot{\theta_{i}} \hat{\textbf{z}} &=& \bm{\tau}_{i} + \bm{\tau}^{\bm{B}}_{i} + \sigma \xi_{i}(t)\hat{\textbf{z}}\label{eqMotion2},
\end{eqnarray}
where $\textbf{r}_{i} = x_{i}\hat{\textbf{x}} + y_{i}\hat{\textbf{y}}$ is the position vector of particle $i$ and $\theta_{i}$ is the angle between the vector $\bm{\mu}_{i}$ and the $x$ axis. $\bm{\tau}_{i}$ and $\bm{\tau}^{\bm{B}}_{i}$ are the torque due to the magnetic field created on particle $i$ by all other particles and the torque created by the external magnetic field $\bm{B}$, respectively. A similar set of equations (\ref{eqMotion1})-(\ref{eqMotion2}) was recently used in Ref.~\cite{LucenaPREMag}, and therefore we report only on the results related to the presence of the modulation in the $x$ direction \cite{Herrera,PREJC}
\begin{equation}\label{EquationMod}
V_{\text{mod}}(x_{i}) = V_{0} \cos\left(\frac{2 \pi x_{i}}{L}\right).
\end{equation}
The parabolic transversal confinement is given by 
\begin{equation}\label{EquationParabolic}
V_{\text{conf}}(y_{i}) = \frac{1}{2}m\omega^{2}y^{2}_{i},
\end{equation}
where $\omega$ is the strength of the confinement (frequency) and $m$ is the mass of the identical particles. Furthermore, the pair interaction potential $V^{\text{int}}_{ij}$ is given by
\begin{equation}\label{eqIntPotential}
V^{\text{int}}_{ij} = \frac{\mu_{0}}{4\pi} \left[ \frac{\bm{\mu}_{i} \cdot \bm{\mu}_{j}}{|\textbf{r}_{ij}|^{3}} - \frac{3(\bm{\mu}_{i} \cdot \textbf{r}_{ij})(\bm{\mu}_{j} \cdot \textbf{r}_{ij})}{|\textbf{r}_{ij}|^{5}} \right] + 4 \varepsilon \left(\frac{\sigma}{|\textbf{r}_{ij}|}\right)^{12},
\end{equation}
where $\mu_{0}$ is the medium permeability, $\textbf{r}_{ij}$ is the inter-particle separation vector between a pair of particles $i$ and $j$ and $\varepsilon$ is an energy parameter in order to prevent particles from coalescing into a single point.

Following previous works \cite{LucenaPREMag,Herrera,PREJC}, we use an Ermak-type algorithm \cite{Ermak} to integrate equations (\ref{eqMotion1})-(\ref{eqMotion2}). The simulations were performed with fixed parameters: $\Delta t = 1.0 \times 10^{-4}(\gamma\sigma^{2}/k_{B}T)$, $\mu = 1.0\sqrt{4 \pi k_{B}T\sigma^{3}/\mu_{0}}$ and $B = 100\sqrt{k_{B}T\mu_{0}/4\pi\sigma^{3}}$. We choose $\varepsilon = k_{B}T$ as unit of energy, $\sigma$ as unit of distance and time is measured in units of $t_{B} = \gamma\sigma^{2}/k_{B}T$. Finally, the stochastic white noise $\bm{\xi}_{i}(t)$ is simulated using the Box-M\"{u}ller transformation technique \cite{BoxMuller} and in all the results presented in this work, the error bars in the plots are smaller than the symbol size. Similarly to our previous paper \cite{LucenaPREMag}, hydrodynamic interactions (HI) are not taken into account. Such interactions may have an impact on the diffusion properties (and in general on the dynamical properties, see, e.g. Ref.~\cite{SabineHYDRO}) for the case of highly concentrated colloidal suspensions, which are not considered in the present work.

%As a final remark, typical values for real (experimental) ferromagnetic colloids are $\sigma = 7$nm and magnetic moment $\mu = 6.0 \times 10^{-20}$Am$^{2}$ \cite{DuboisJChemPhys,PTurq}.

In order to study diffusion we calculate the mean-square displacement (MSD) $W(t)$, defined as \cite{SaintJeanPRE}
\begin{equation}\label{MSDsystem}
W(t) = \left\langle N^{-1} \sum_{i=1}^{N} |\textbf{r}_{i}(t) - \textbf{r}_{i}(0)|^{2} \right\rangle,
\end{equation}
where we use a typical value of $N=300-900$ particles, $t$ is the time and $\langle \ldots \rangle$ is an average over different time origins during the simulation \cite{FrenkelUNDER}. This equation can be split in two terms, namely $W_{x}(t)$ and $W_{y}(t)$, where the first refers to the mean-square displacement in the $x$ direction and the latter refers to the mean-square displacement in the $y$ direction.

The system is tuned by three parameters, namely (i) the linear density, $\rho = N/L_{x}$ where $L_{x}$ is the size of the computational unit cell in the $x$ direction and $N$ is the total number of particles, (ii) the angle $\phi$ of the external magnetic field, and (iii) the strength $V_{0}$ of the external modulation in the $x$ direction. Note that since we are using periodic boundary conditions in the $x$ direction, we have to guarantee  the continuity of the external modulation at the borders of the computational unit cell. This is achieved by introducing the relation
\begin{equation}
L_{x} = nL,\label{eqLxL}
\end{equation}
where $n \in Z^{+}$ and it represents the number of minima (or maxima) of the external modulation within the computational unit cell.

\section{Linear density $\rho = 0.5 \sigma^{-1}$}\label{MSDSecDensity}

\subsection{Case $\omega = 1.0 \sqrt{2k_{B}T/m\sigma^{2}}$}\label{dens05w1}
\begin{figure}[ht]
\begin{center}
%\hspace*{-0.2cm}
\includegraphics[width=8.6cm]{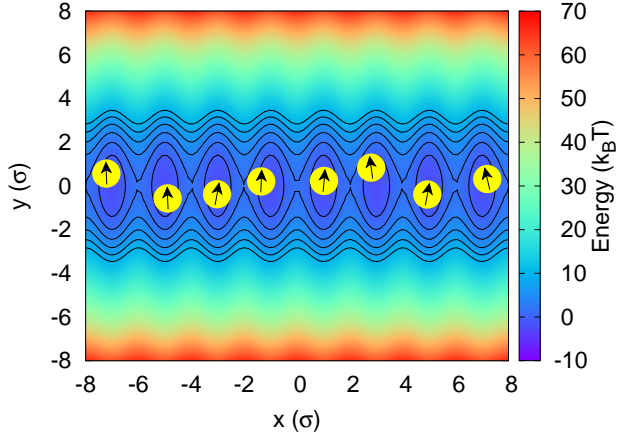}
%\vspace*{-1.0cm}
\end{center}
\caption{(Color online) Snapshot of the configuration of the system for $V_{0}/k_{B}T = 2.0$. The particles are represented by yellow circles where the black arrows indicate the direction of the dipoles. The contour plot of the potential $V_{\text{mod}}(x) + V_{\text{conf}}(y)$ is also shown. The linear density is $\rho = 0.5\sigma^{-1}$ and the transversal confinement strength is $\omega = 1.0 \sqrt{2k_{B}T/m\sigma^{2}}$.}\label{ConfigVo2w1}
\end{figure}

\begin{figure}[ht]
\begin{center}
%\hspace*{-0.2cm}
\includegraphics[width=8.6cm]{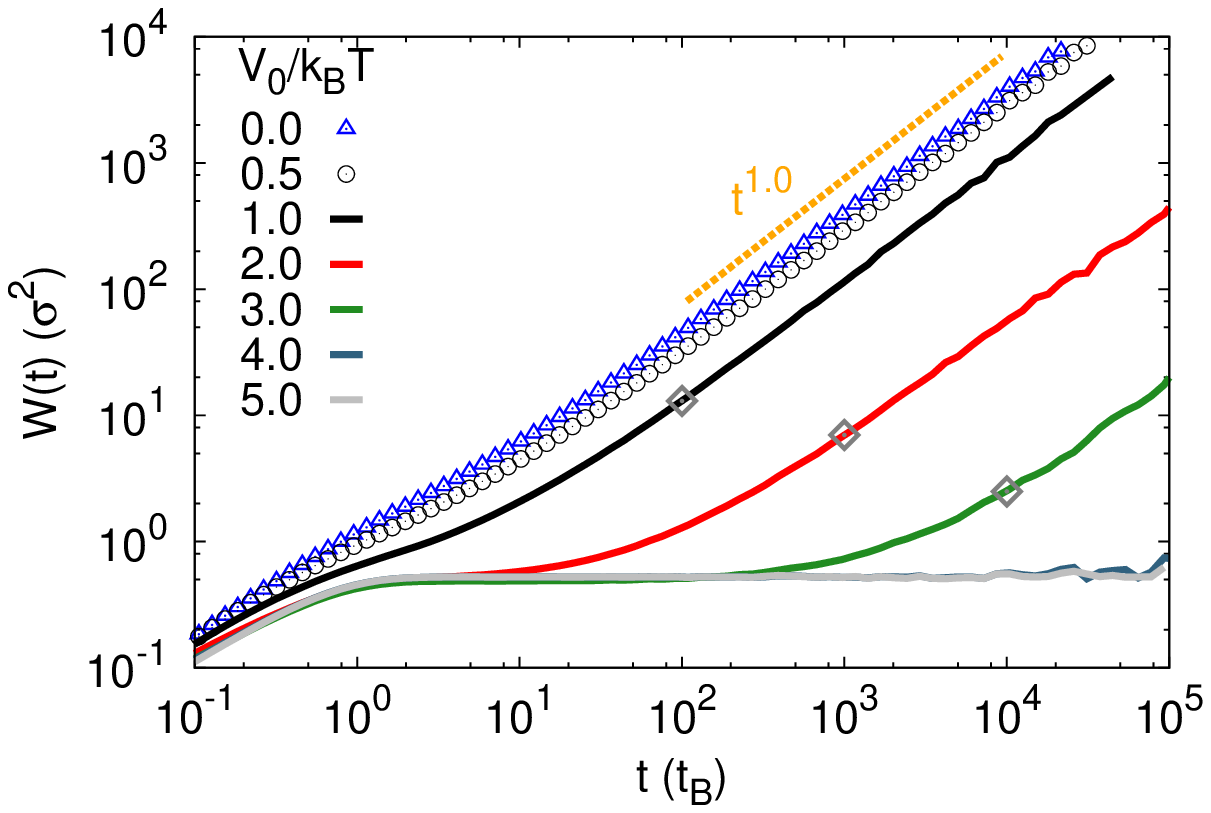}
%\vspace*{-1.2cm}
\end{center}
\caption{(Color online) Log-log plot of the mean-square displacement $W(t)$ as a function of time $t$ for different values of the ratio $V_{0}/k_{B}T$. The yellow dotted line is a guide to the eye. The open diamonds indicate approximately the time scale ($t_{N}$) where the normal diffusive regime, i.e. $W(t) \propto t$,  is recovered. The transversal confinement strength is $\omega = 1.0 \sqrt{2k_{B}T/m\sigma^{2}}$ and the linear density is $\rho = 0.5 \sigma^{-1}$.}\label{MSD_w1}
\end{figure}

In this section, we set the transversal confinement parameter $\omega = 1.0 \sqrt{2k_{B}T/m\sigma^{2}}$ and $\phi=90^{o}$. A snapshot of the configuration of the system together with the contour plot of the periodic modulation and transversal confinement is shown in Fig~\ref{ConfigVo2w1}. The mean-square displacement $W(t)$ [Eq.~(\ref{MSDsystem})] for different values of $V_{0}/k_{B}T$ is shown in Fig.~\ref{MSD_w1}. Note that for all the values of $V_{0}/k_{B}T$, except for $4.0$ and $5.0$, $W(t)$ exhibits a linear dependence on time $t$ for large time scales:
\begin{equation} \label{MSDsimw1}
\lim_{t \gg t_{N}}W(t) = D_{s}t, 
\end{equation}
where $D_{s}$ is the self-diffusion coefficient and $t_{N}$ (indicated by gray open diamonds) is the time scale at which this normal diffusion regime is recovered. Note that since the system is coupled to a heat bath ($k_{B}T$), the normal diffusion regime should be recovered for any value of the ratio $V_{0}/k_{B}T$, with the condition that $t_{N} \rightarrow \infty$ for $V_{0}/k_{B}T \rightarrow \infty$. In other words, this means that the intermediate regime (i.e., where $W(t)$ exhibits a slower-than-linear dependence on time or $W(t) = \text{const}$) extends over a larger time interval for larger values of $V_{0}/k_{B}T$. This intermediate regime is generally associated with a ``cage'' effect, which in our case is represented by the localization of particles in the potential minima. A similar effect was found previously in simulations on monodisperse glassy systems \cite{KobAndPRL} and Lennard-Jones binary mixtures \cite{SzamelPRE}. However, in these works, the caging effect was not induced by an external modulation but rather by many-body effects related to the specificities of their system.

\begin{figure}[ht]
\begin{center}
%\hspace*{-0.2cm}
\includegraphics[width=8.6cm]{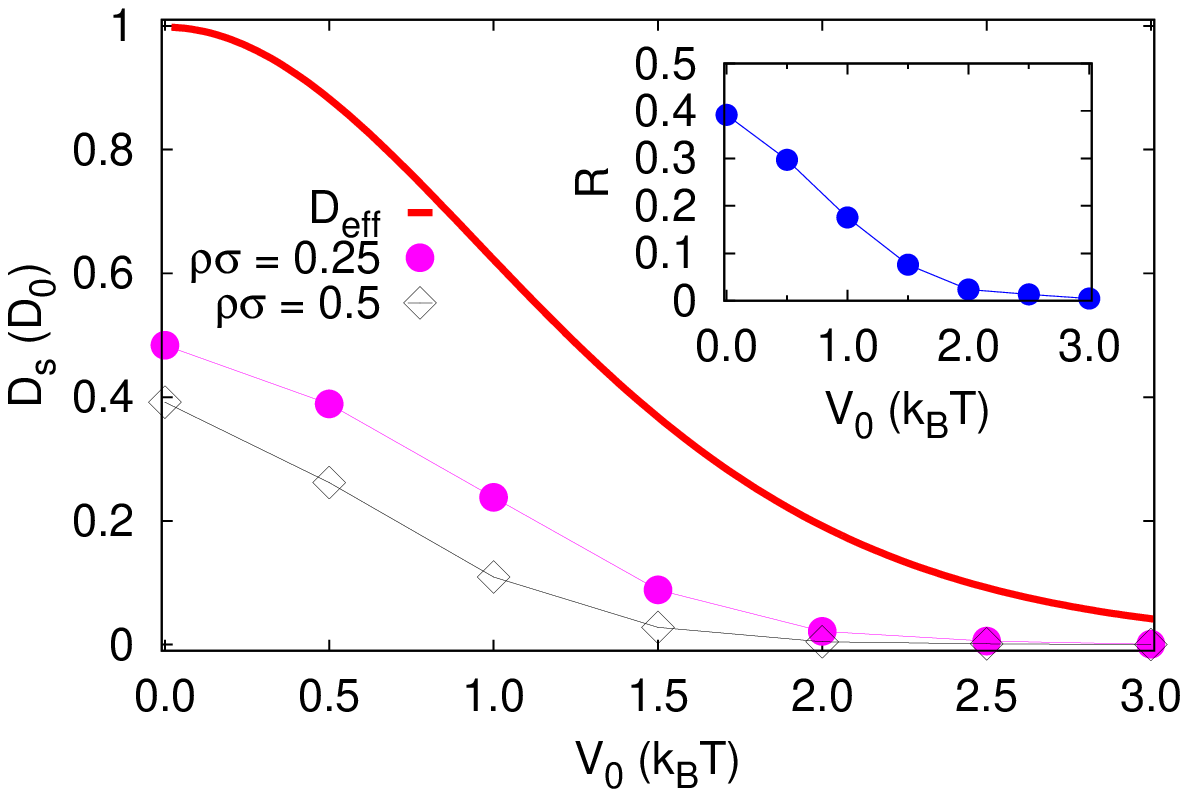}
%\vspace*{-2.0cm}
\end{center}
\caption{(Color online) Long-time self-diffusion coefficient $D_{s}/D_{0}$ as a function of the ratio $V_{0}/k_{B}T$, for different linear densities $\rho$. The effective self-diffusion coefficient $D_{\text{eff}}/D_{0}$ [Eq.~(\ref{Deff})] as a function of $V_{0}/k_{B}T$ for a single particle is also shown (solid red curve) for comparison. The inset shows the ratio $R = D_{s}/D_{\text{eff}}$ as a function of $V_{0}/k_{B}T$ for the case $\rho = 0.5 \sigma^{-1}$.}\label{Dsim}
\end{figure}

From these results, we also note that, as expected from previous section, the self-diffusion coefficient depends on the ratio $V_{0}/k_{B}T$. This dependence is shown in Fig.~\ref{Dsim}, where $D_{s}$ decreases with increasing $V_{0}/k_{B}T$. $D_{s}$ is obtained by fitting our data with Eq.~(\ref{MSDsimw1}). Note that even though the behaviour of $D_{s}$ as a function of the ratio $V_{0}/k_{B}T$ is qualitatively similar to $D_{\text{eff}}(V_{0}/k_{B}T)$ for a single-particle, it is clear that $D_{s} < D_{\text{eff}}$. This difference between $D_{s}$ and $D_{\text{eff}}$ is due to correlations between the particles, which now couples the movement of the dipoles through the interaction potential. We estimate this difference by calculating the ratio $R = D_{s}/D_{\text{eff}}$ which is shown in the inset of Fig.~\ref{Dsim}. Note that $R$ drops to zero as $V_{0}/k_{B}T$ increases. This means that in both cases, i.e., for single-particle and for interacting particles, the self-diffusion coefficient goes to a value very close to zero (but does not vanish completely, see Ref.~\cite{StefanEgelArx}, Sec.~IIIB) as $V_{0}/k_{B}T$ increases. Therefore, there is no diffusion until temperature is sufficiently high to allow the escape of the particles from the potential wells \cite{BaoAi}. The effect of the linear density $\rho$ on the self-diffusion coefficient, $D_{s}$, will be discussed in Sec.~\ref{linDens}.

\subsection{Case $\omega = 10.0 \sqrt{2k_{B}T/m\sigma^{2}}$}
In the case where the transversal confinement potential is increased, the fluctuations of the particles in the $y$ direction becomes smaller. This effect of confinement brings the system into the single-file (SF) regime, which means that particles cannot bypass each other \cite{LucenaPRE2012}. This special geometric constraint leads to a phenomenon called single-file diffusion (SFD), in which one of the most striking feature is that the long-time mean-square displacement $W_{x}(t)$ of a tagged particle along the unconfined direction (in our case, the $x$ direction) displays typical sub-diffusive motion with
\begin{equation}\label{msdSFD}
\lim_{t \gg t_{c}}W_{x}(t) = Ft^{0.5},
\end{equation}
where $F$ is the so-called single-file diffusion mobility and $t_{c}$ is a characteristic relaxation time of the system. In particular, $F$ and $t_{c}$ depend on the specifics of the system \cite{KrapArxiv}. Wei \textit{et al.} \cite{WeiScience} showed experimentally that for a repulsive inter-particle interaction potential, $t_{c}$ decreases with increasing strenght of the interaction potential. This can be understood from the fact that an increase in the interaction leads to an increase in the collision rate between the particles \cite{BeijerenPRB}. Nelissen \textit{et al.} \cite{NelissenEPL} recently showed that when the inter-particle interaction is comparable to the viscosity (damping), an intermediate ``under'' single-file diffusion regime, i.e. $W_{x}(t) \propto t^{\alpha}$ (with $\alpha < 0.5$), is also observed. Such a behaviour was also found in experiments with millimetre metallic balls \cite{SaintJeanPRE2006} and in numerical simulations \cite{SlavaPREDimitri} taking into account spatial correlated noises.

\begin{figure}[ht]
\begin{center}
%\hspace*{-0.2cm}
\includegraphics[width=8.6cm]{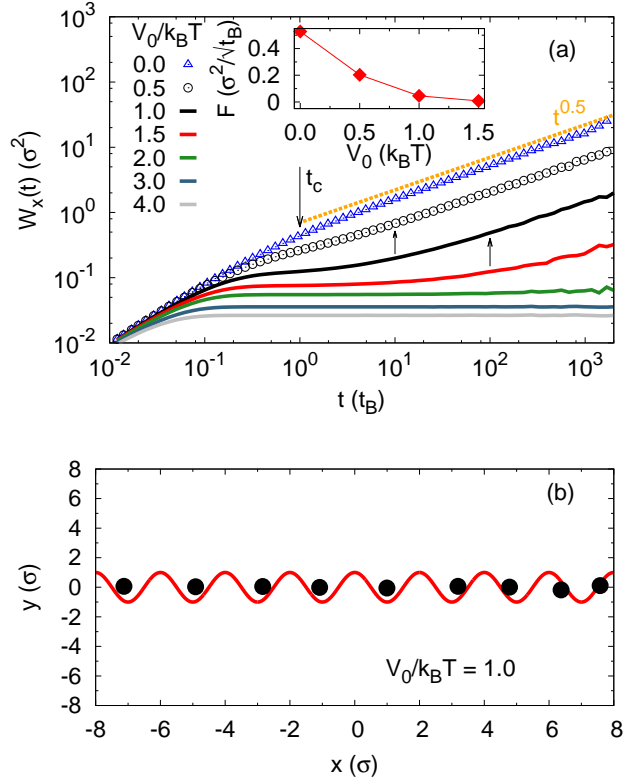}
%\vspace*{-1.0cm}
\end{center}
\caption{(Color online) (a) Log-log plot of the mean-square displacement in the $x$ direction, $W_{x}(t)$, as a function of time $t$ for different values of the ratio $V_{0}/k_{B}T$. The yellow dotted line is a guide to the eye. The transversal confinement strength is $\omega = 10.0 \sqrt{2k_{B}T/m\sigma^{2}}$ and the linear density is $\rho = 0.5\sigma^{-1}$. Vertical black arrows indicate the relaxation time $t_{c}$. Inset: Single-file diffusion mobility $F$, obtained from the relation (\ref{msdSFD}), as a function of $V_{0}/k_{B}T$. (b) Snapshot of the configuration of particles (black dots) for $V_{0}/k_{B}T = 1.0$. The modulation $V_{\text{mod}}(x)$ is plotted as the solid red curve.}\label{MSD_w10}
\end{figure}

In our specific case, the modulation in the $x$ direction adds an additional restriction to the movement of the particles. The effect of $V_{0}/k_{B}T$ on the mean-square displacement $W_{x}(t)$ is shown in Fig.~\ref{MSD_w10}(a). Two effects are noticed here: First, the relaxation time $t_{c}$ increases with increasing ratio $V_{0}/k_{B}T$, which means that for higher values of this ratio a longer time is needed for a particle to feel the presence of its neighboring particles. Once this time scale is reached, the sub-diffusive law (\ref{msdSFD}) is recovered due to the interaction with its neighbors. Second, the mobility factor $F$ decreases with increasing $V_{0}/k_{B}T$ [cf. inset of Fig.~\ref{MSD_w10}(a)], which results from the restriction of the motion in the $x$ direction, as stated above. Note that for $V_{0}/k_{B}T > 0.0$, the system exhibits an intermediate regime where $W_{x}(t) \propto t^{\alpha}$, with $\alpha < 0.5$ before it reaches the SFD regime. This intermediate regime extends to larger times scales as the ratio $V_{0}/k_{B}T$ increases.

%We fit the points $2F$ vs. $V_{0}$ obtained from the relation () with an exponential decay and find that $2F = F_{0}\exp\{-(V_{0}/V_{m})\}$, with $F_{0} = (0.531 \pm 0.021) \sigma^{2}/\sqrt{t_{B}}$ and $V_{m} = (0.473 \pm 0.043) k_{B}T$.

\section{Effect of linear density on diffusion}\label{linDens}
\begin{figure}[ht]
\begin{center}
%\hspace*{-0.2cm}
\includegraphics[width=8.6cm]{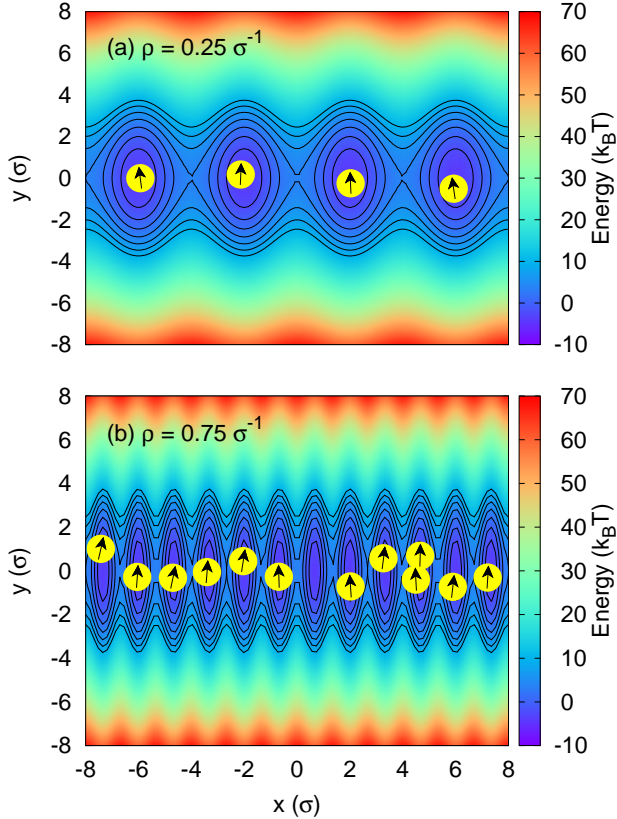}
%\vspace*{-1.2cm}
\end{center}
\caption{(Color online) The same as Fig.~\ref{ConfigVo2w1} but now for $V_{0}/k_{B}T = 4.0$. Linear density is (a) $\rho = 0.25\sigma^{-1}$, and (b) $\rho = 0.75\sigma^{-1}$. For both cases, the transversal confinement strength is $\omega = 1.0 \sqrt{2k_{B}T/m\sigma^{2}}$ and the commensurability factor is $p=1$.}\label{FigDens0251}
\end{figure}

\begin{figure}[ht]
\begin{center}
%\hspace*{-0.2cm}
\includegraphics[width=8.6cm]{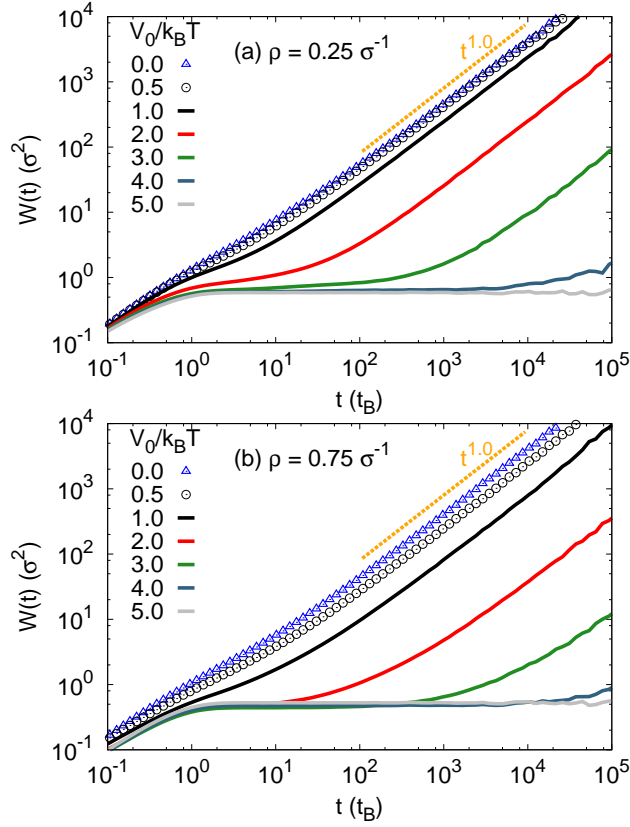}
%\vspace*{-1.2cm}
\end{center}
\caption{(Color online) The same as Fig.~\ref{MSD_w1} but now for density (a) $\rho=0.25\sigma^{-1}$, and (b) $\rho=0.75\sigma^{-1}$. The transversal confinement strength is $\omega = 1.0 \sqrt{2k_{B}T/m\sigma^{2}}$ and the commensurability factor is $p=1$.}\label{MSDDens0251}
\end{figure}

In order to investigate the effect of the linear density $\rho$ on the diffusion, we introduce a commensurability factor $p \equiv N/n$, where $N$ is the total number of particles in the computational unit cell and $n$ is the total number of minima (or maxima) of the external periodic modulation along the $x$ direction. Using relation (\ref{eqLxL}) and the definition for the linear density, we may write the following condition
\begin{equation}
p \equiv \frac{N}{n} = \rho L.
\end{equation}
We start by considering the simplest case ($p=1$), i.e., where there is one particle per potential well. In this section we analyse the system for three different densities, namely $\rho\sigma = 0.25, 0.50, 0.75$. Also, we fix $\omega = 1.0 \sqrt{2k_{B}T/m\sigma^{2}}$ and $\phi = 90^{o}$. In Figs.~\ref{FigDens0251}(a)-(b) we show snapshots of the configuration of the system for $\rho=0.25\sigma^{-1}$ and $\rho=0.75\sigma^{-1}$, respectively. The mean-square displacement $W(t)$ for different values of $\rho$ is shown in Figs.~\ref{MSDDens0251}(a)-(b).

The main effect of different densities on $D_{s}$ is shown in Fig.~\ref{Dsim}. The solid curve is the single-particle case discussed in Sec.~\ref{singlePart}, which corresponds to the limiting case of very dilute systems, i.e., very low densities. As the density increases ($\rho = 0.25\sigma^{-1}$ and $\rho = 0.5\sigma^{-1}$), the self-diffusion coefficient $D_{s}$ decreases. This effect is related to the coupling between the particles due to the inter-particle interaction potential. For the case of very high densities, the interaction energy is stronger and diffusion should be partially suppressed, i.e., $D_{s} \approx 0$ for all values of $V_{0}/k_{B}T$. Note that since the system is coupled to a heat bath, the diffusion coefficient is \textit{not exactly} zero but goes to a very small value.

\section{Effect of commensurability factor}\label{CommSection}
\begin{figure*}[ht]
\begin{center}
%\hspace*{-0.2cm}
\includegraphics[width=17.2cm]{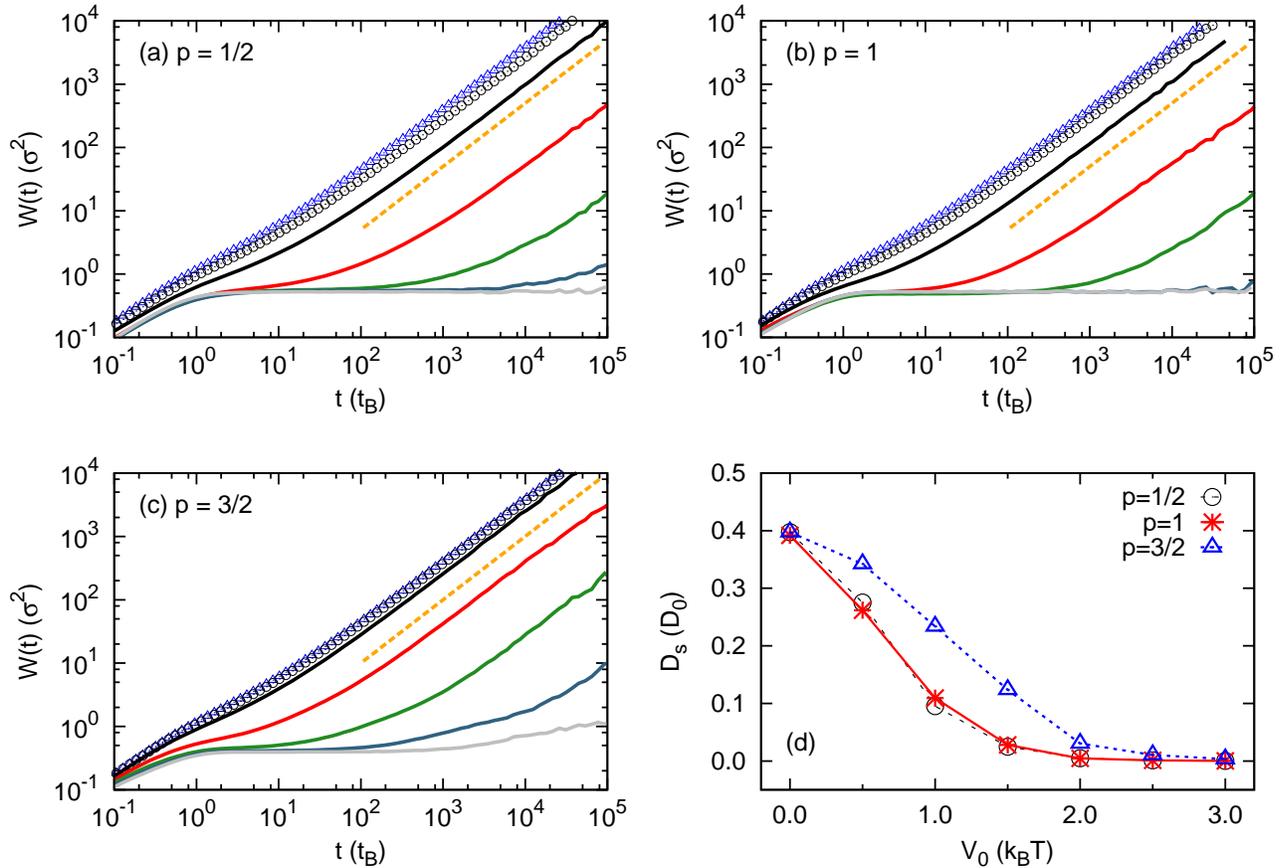}
%\vspace*{-0.8cm}
\end{center}
\caption{(Color online) (a)-(c) Log-log plot of the mean-square displacement $W(t)$, as a function of time $t$ for different values of the ratio $V_{0}/k_{B}T$. The yellow dotted line has a slope of 1 and is a guide to the eye. The transversal confinement strength is $\omega = 1.0 \sqrt{2k_{B}T/m\sigma^{2}}$ and the linear density is $\rho = 0.5\sigma^{-1}$. Color code is the same as in Fig.~\ref{MSDDens0251}. (d) Long-time self-diffusion coefficient, $D_{s}$, as a function of $V_{0}/k_{B}T$ for different values of the commensurability factor $p$.}\label{MSDChangep}
\end{figure*}

We further investigate the effect of the commensurability factor $p$ on the self-diffusion coefficient. In this section, we fix the linear density to $\rho = 0.5 \sigma^{-1}$ and vary $p$, where we choose two half-integer values ($p=1/2$ and $p=3/2$) and compare these results with the case of previous section ($p=1$). The effect of $p$ on the mean-square displacement $W(t)$ is shown in Figs.~\ref{MSDChangep}(a)-(c). Note that for all cases, the system exhibits an intermediate regime of diffusion where $W(t)$ shows a slower-than-linear dependence on time or $W(t) = \text{const}$ before the long-time normal diffusion regime sets in [Eq.~(\ref{MSDsimw1})]. See discussion in Sec.~\ref{dens05w1}.

\begin{figure}[ht]
\begin{center}
%\hspace*{0.0cm}
\includegraphics[width=6.5cm]{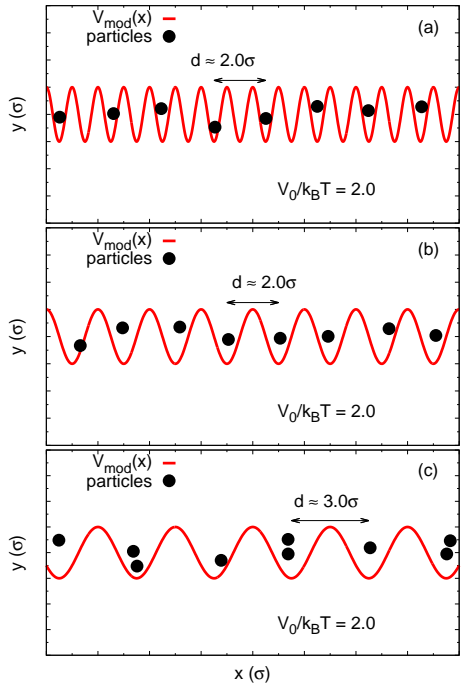}
%\vspace*{-0.6cm}
\end{center}
\caption{(Color online) Snapshot of the configuration of the system for different values of the commensurability factor $p=$ (a) $1/2$, (b) $1$ and (c) $3/2$. For all cases, the strenght of the $x$ direction modulation is $V_{0}/k_{B}T = 2.0$. Note that $L$ changes according to the value of $p$.}\label{ConfModP}
\end{figure}

An interesting effect of the commensurability factor $p$ on diffusion can be observed in Fig.~\ref{MSDChangep}(d). For $V_{0}/k_{B}T = 0.0$, the self-diffusion coefficient, $D_{s}$, is the same for all the cases (i.e., $p=1/2,1,3/2$). This is due to the fact that in the absence of the external modulation, the system is regulated only by the linear density (in this case $\rho = 0.5\sigma^{-1}$). Therefore, the average distance between neighbour particles is the same. On the other hand, for sufficiently large values of $V_{0}/k_{B}T = 3.0$, the trapping of particles in the wells suppresses the diffusion, and again the self-diffusion coefficient $D_{s}$ is of the same order (close to zero) for all the cases. However, the effect of $p$ on $D_{s}$ is more pronounced for intermediate values of $V_{0}/k_{B}T = 0.5 - 2.0$. This effect is explained as follows. First, note that $p=1/2$ and $p=1$ have both very similar behaviours, i.e., $D_{s}$ curve as a function of $V_{0}/k_{B}T$. From the definition of $p$, we have that for $p=1/2 \rightarrow L=1.0\sigma$ and $p=1 \rightarrow L=2.0\sigma$. In practice, this means that the neighbour inter-particle \textit{average} distance is the same for both cases, i.e., $d \approx 2.0\sigma$ [cf. Figs.~\ref{ConfModP}(a)-(b)]. For $p=3/2$ (which means $3$ particles per $2$ potential wells, on average), the distance between particles in neighbouring wells is larger, $d \approx 3.0\sigma$, which results in a larger self-diffusion coefficient. Interestingly, this case can be thought as a binary system, where one of the wells has one ``big'' particle formed by two dipoles and the other well has only one particle. Note that for all cases, $D_{s}$ decreases with increasing $V_{0}/k_{B}T$, although for $p=3/2$ this decrease is slower compared to the other cases ($p=1$ and $p=1/2$).

\section{Anisotropic diffusion and transversal sub-diffusion}\label{AnisoSFD}

\subsection{Two particles per potential well}
\begin{figure}[ht]
\begin{center}
%\hspace*{-0.2cm}
\includegraphics[width=8.6cm]{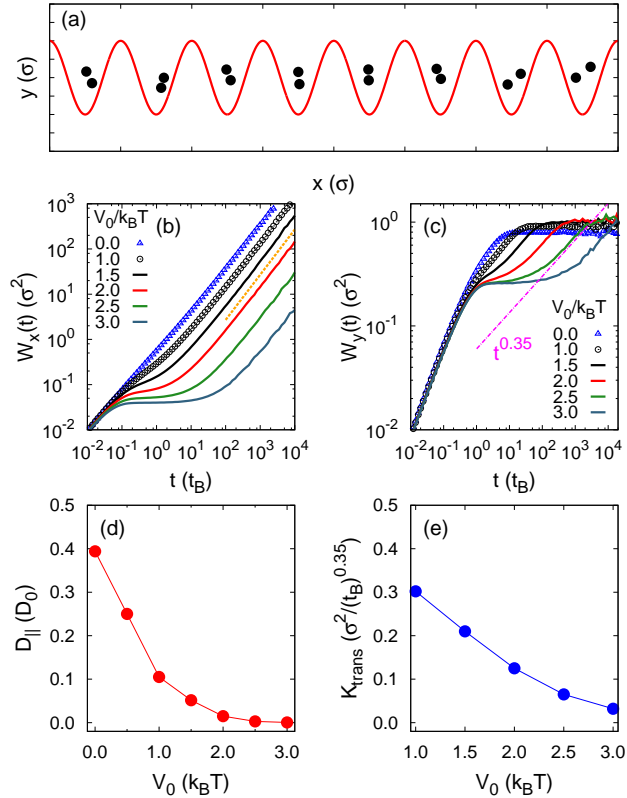}
%\vspace*{-1.0cm}
\end{center}
\caption{(Color online) (a) Snapshot of the configuration of particles (black dots) for $V_{0}/k_{B}T = 3.0$. The modulation $V_{\text{mod}}(x)$ is plotted as the solid red curve. (b), (c) Log-log plot of the MSD as a function of time $t$ in the parallel and transversal direction, respectively, for different values of $V_{0}/k_{B}T$. The dotted yellow line has a slope of 1, the magenta dotted-dashed line has a slope of 0.35 and both are guide to the eye. (d) Parallel self-diffusion coefficient $D_{||}$ and (e) anomalous transversal diffusion coefficient $K_{\text{trans}}$, both as a function of $V_{0}/k_{B}T$. Parameters of the simulation are $p=2$, $\rho=1.0\sigma^{-1}$ and $\omega = 1.0 \sqrt{2k_{B}T/m\sigma^{2}}$.}\label{MSD_wx_wy}
\end{figure}

The competition between the external potentials in the $x$ and $y$ directions (i.e. the modulation [Eq.~(\ref{EquationMod})] and the parabolic potential [Eq.~(\ref{EquationParabolic})], respectively) leads to an anisotropic diffusion process, i.e., $W_{x}(t) \neq W_{y}(t)$ \cite{NoteMSDXandY}. In this section we analyse the effect of the ratio $V_{0}/k_{B}T$ on both the parallel ($x$ direction) and transversal ($y$ direction) diffusion independently. For this case, the simulation parameters are $p=2$, $\rho=1.0\sigma^{-1}$ and $\omega = 1.0 \sqrt{2k_{B}T/m\sigma^{2}}$, which allows the accommodation of two particles per potential well on \textit{average} [cf. Fig.~\ref{MSD_wx_wy}(a)]. As a representative example, we show in Figs.~\ref{MSD_wx_wy}(b)-(c) the MSD in the parallel and transversal direction, respectively, for different values of $V_{0}/k_{B}T$. Note that the diffusion in the parallel direction is very different from the transversal direction, which is a direct effect of the anisotropy of space, i.e., the competition between periodic modulation in the $x$ direction and the parabolic confinement in the $y$ direction.

In the $x$ direction (parallel diffusion), the MSD exhibits [cf. Fig.~\ref{MSD_wx_wy}(b)] a short-time normal diffusion behaviour for $t < t_{B}$, which is followed by a saturation regime due to the periodic modulation. Finally, for $t \gg t_{B}$, the long-time normal diffusion regime is recovered, with
\begin{equation}
\lim_{t \gg t_{B}} W_{x}(t) = D_{||}t,
\end{equation}
where $D_{||}$ is the parallel self-diffusion coefficient. The dependence of $D_{||}$ on $V_{0}/k_{B}T$ is shown in Fig.~\ref{MSD_wx_wy}(d), and as expected it decreases with increasing $V_{0}/k_{B}T$.

On the other hand, in the $y$ direction (transversal diffusion), the MSD exhibits [cf. Fig.~\ref{MSD_wx_wy}(c)] a very different behaviour. The initial short-time normal diffusion is also present. However, for intermediate time scales $t_{B} < t < t_{\text{sat}}$ the system exhibits a sub-diffusive regime with a non-linear time-dependence of the form
\begin{equation}
W_{y}(t) = K_{\text{trans}}t^{\alpha},
\end{equation}
where $K_{\text{trans}}$ is the anomalous transversal diffusion coefficient \cite{BarkaiPhysToday} and $t_{\text{sat}}$ is a saturation time scale in which the diffusion is suppressed due to the finite-size in the $y$ direction. Note that $\alpha<0.5$ and thus a smaller power-law behaviour as compared to the single-file diffusion (SFD) case. In Fig.~\ref{MSD_wx_wy}(c) we show this intermediate regime and find $\alpha \approx 0.35$. Finally, both $K_{\text{trans}}$ [cf. Fig.~\ref{MSD_wx_wy}(e)] and $t_{\text{sat}}$ depends on the periodic modulation strength $V_{0}/k_{B}T$, which is a measure of a type of ``effective'' confinement in the $x$ direction. This indicates that the periodic modulation in the parallel direction affects directly the diffusion process in the transversal direction. A similar transversal sub-diffusive behaviour was recently found and analysed by Delfau \textit{et al.} \cite{SJeanTrans1} in a quasi-one-dimensional system of interacting particles in a thermal bath. Note that even for the case of two particles in a single-file condition, Ambj\"{o}rnsson and Silbey \cite{AmbJChemPhys} showed that the long-time SFD regime should appear. The fact that we cannot observe the same behaviour here is because of a competition between the inter-particle interaction and the finite-size effect in the transversal direction. Note that our discussion is only valid for an \textit{intermediate regime} (ITR) of sub-diffusion, as discussed previously in Refs.~\cite{LucenaPREMag,LucenaPRE2012} and references therein.

\subsection{Four particles per potential well}
\begin{figure}[ht]
\begin{center}
%\hspace*{-0.2cm}
\includegraphics[width=8.6cm]{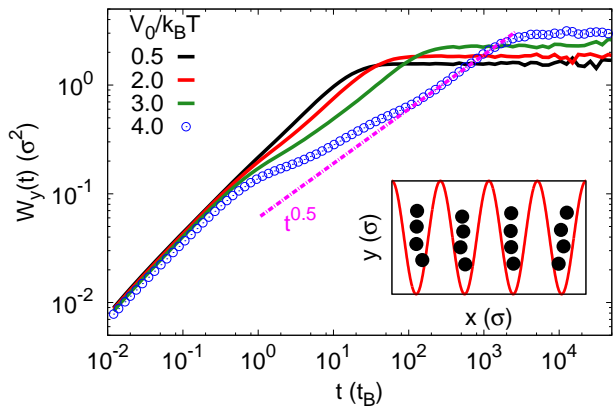}
%\vspace*{-1.2cm}
\end{center}
\caption{(Color online) Log-log plot of the transversal MSD $W_{y}(t)$ as a function of time $t$, for different values of $V_{0}/k_{B}T$. The magenta dotted-dashed line has a slope of 0.5 and is a guide to the eye. Inset: snapshot of the configuration of particles (black dots) for $V_{0}/k_{B}T = 4.0$. The modulation $V_{\text{mod}}(x)$ is plotted as the solid red curve. Parameters of the simulation are $p=4$, $\rho=2.0\sigma^{-1}$ and $\omega = 1.0 \sqrt{2k_{B}T/m\sigma^{2}}$.}\label{MSDy_P4}
\end{figure}

In this section we analyse the transversal diffusion mechanism for $p=4$, which gives four particles per potential well. As in the previous section, we calculate the transversal MSD $W_{y}(t)$ as a function of the strength of the external periodic modulation $V_{0}/k_{B}T$, and the results are shown in Fig.~\ref{MSDy_P4}.

For the case of weak external modulation (e.g. $V_{0}/k_{B}T = 0.5$), the initial short-time linear MSD [$W_{y}(t) \propto t$] is followed by a saturation regime due to the finite size of the system in the transversal direction. With the increase of the external modulation, an intermediate sub-diffusive regime takes place before the onset of the saturation regime [cf. Fig.~\ref{MSDy_P4}, $V_{0}/k_{B}T = 4.0$]. This is explained by the formation of a chain of particles along the transversal direction [cf. inset of Fig.~\ref{MSDy_P4}]. Note that, as opposed to the previous section, where $W_{y}(t) \propto t^{0.35}$, in this case the MSD presents a clear SFD scaling, i.e., $W_{y}(t) \propto t^{0.5}$.

These results indicate that even though the chain of particles is relatively small, the correlations between particles is sufficiently strong \cite{SJeanTrans1} to induce an intermediate single-file diffusion regime.

\section{Conclusions}\label{ConcDiss}
We studied the diffusive properties of a system of interacting magnetic dipoles in the presence of a modulated (corrugated) channel along the $x$ direction and confined in the $y$ direction by a parabolic confinement potential. In order to study the diffusion of the system, we used Brownian dynamics simulations. The analysis of the mean-square displacement $W(t)$ showed that the system exhibits different regimes of diffusion depending on the external parameters (i.e. external modulation, magnetic field) that regulate the particle dynamics. In principle, this system could be realised experimentally using optical tweezer traps and our results could be verified by, e.g., a microscopy imaging technique to track individual particles' trajectories \cite{EgelhaafSM2012,JGoreePRL}.

We characterized the dynamics of the system for several parameters, namely the linear density $\rho$, the commensurability factor $p$ and the strength of the external periodic modulation $V_{0}/k_{B}T$. Our main results are summarized as follows: (i) the self-diffusion coefficient $D_{s}$ is modified by the inter-particle interaction potential as compared to the case of a single-particle diffusing in a periodic potential landscape. The difference increases with the linear density of particles; (ii) the effect of the commensurability factor $p$ on the self-diffusion coefficient $D_{s}$ is pronounced for the case of a semi-integer commensurability factor (as an example we considered $p=3/2$). The system turns into an effective ``artificial'' binary system, with the presence of a ``big'' particle formed by two dipoles in a potential well and a single particle in a neighbour potential well; (iii) the presence of the external modulation affects the diffusion of the magnetic dipoles as compared to the case where there is no modulation (cf. Ref.~\cite{LucenaPREMag}); for instance, we found that a transversal sub-diffusive regime, including SFD, can be induced depending on the value of the external modulation $V_{0}/k_{B}T$ and on the commensurability factor $p$.

\section*{ACKNOWLEDGMENTS}
This work was supported by CNPq, CAPES, FUNCAP (Pronex grant), the Flemish Science Foundation (FWO-Vl), the collaborative program CNPq - FWO-Vl, and the Brazilian program Science Without Borders (CsF). D.~Lucena acknowledges fruitful discussions with W.~A.~Mu\~{n}oz, V.~F.~Becerra, E.~C.~Eu\'{a}n-D\'{i}az and M.~R.~Masir.


\begin{references}
\bibitem{NoteMSDXandY}
Note that the mean-square displacement (MSD) in the $x$ and $y$ direction are calculated similarly to Eq.~(\ref{MSDsystem}), where $W_{\lambda}(t) = \langle N^{-1}\sum_{i=1}^{N} [\lambda_{i}(t) - \lambda_{i}(0)]^{2} \rangle$ and $\lambda_{i}$ is the $x$ or $y$ coordinate of each $i$th particle.
\end{references}
\end{document}